\documentclass[twocolumn,groupedaddress,superscriptaddress]{revtex4}

\usepackage[greek,english]{babel}
\usepackage{graphicx}
\usepackage{amsmath,bm}
\usepackage{mathrsfs}
\usepackage{amsfonts}
\usepackage{amssymb}
\usepackage[utf8]{inputenc}
\usepackage{booktabs}
\usepackage{booktabs,amsmath}

\begin{document}

\title{Longitudinal near-field coupling between acoustic resonators grafted onto a waveguide}

\author{Yan-Feng Wang}
\author{Vincent Laude}
\email{vincent.laude@femto-st.fr}
\affiliation{Institut FEMTO-ST, CNRS UMR 6174, Universit\'e de Bourgogne Franche-Comt{\'e}, Besan\c{c}on, F-25030, France}

\begin{abstract}
We investigate longitudinal near-field coupling between acoustic resonators grafted onto a waveguide.
Experiments are performed in the audible range with a simple acoustic system composed of a finite aperiodic sequence of air resonators.
Transmission typically shows a zero around a resonance frequency of a single resonator, as is well known.
When two identical resonators are brought in close proximity, however, we observe that longitudinal near-field coupling strongly influences the acoustic transmission.
When the separation between resonators is increased so that they can be considered in the far field of one another, we further observe the appearance of Fano-like transmission profiles.
We explain this observation by the formation of locally resonant Fabry-Perot interferometers from every pair of resonators.
All experimental results are compared to three-dimensional finite element analysis of the acoustic system.

\end{abstract}


\maketitle

\section{Introduction}
\label{sec1}

Control of acoustic/elastic wave propagation with small-size structures and in the low-frequency range is currently becoming an urgent demand. 
In 2000, Liu \textit{et al.} proposed a sonic crystal exhibiting band gaps though the lattice constant was two orders of magnitude smaller than the relevant sonic wavelength \cite{liuSCIENCE2000}.
Those band gaps were generated owing to local resonances, with little dependence on periodicity.
It was further shown that locally resonant structures can also present negative effective properties at frequencies associated with the local resonance \cite{laiNatMater2011,zhuNatC2014}.
The use of local resonance for band gap engineering and the achievement of other intriguing properties have indeed paved new paths for the control of wave propagation.

Numerous works have been devoted to the design of resonant acoustic elements.
Various configurations have been proposed, including systems with cross-like holes in a solid medium \cite{wang2011JAP}, very soft inclusions in a binary system \cite{wang2004PRL}, split rings inside water \cite{elford2011jasa}, inertial amplifiers \cite{yilmaz2007prb}, lattice structures with suspended masses \cite{ma2013OE}, plates or surfaces with pillars \cite{pennec2008PRB, achaoui2011PRB}, or Helmholtz resonators \cite{fangNatMat2006,lemoult2011PRL}.
The corresponding line of research has broad scientific and practical implications into harnessing the propagation of acoustic waves \cite{husseinAMR2014}.

When resonators are brought at a close distance, typically smaller than the operating wavelength, near-field coupling between them may occur.
In addition to transverse coupling \cite{fuAFM2011}, longitudinal near-field coupling, i.e. coupling between adjacent units in the direction of wave propagation, is for instance employed to tune the response of photonic crystals \cite{liuNatPhoto2009, powellPRB2011}.
In the context of locally resonant phononic or sonic crystals, longitudinal coupling has been shown to introduce deviations from simple models that neglect it \cite{wangJPD2014}.
In this paper, we investigate longitudinal near-field coupling between acoustic resonators.
We specifically consider a very simple and inexpensive system composed of a finite sequence of air resonators grafted onto a waveguide.
Measurements are performed in the audible range.
It is found that longitudinal near-field coupling has very significant effects on the transmission through the waveguide.
Furthermore, as we increase the separation between two identical resonators, spanning the range from the near to the far field, we observe a transition from a collective response of the near-field coupled resonators to a Fano-like transmission profile arising from pairs of resonators forming Fabry-Perot (FP) interferometers.

\begin{figure}[!t]
\centering
\includegraphics{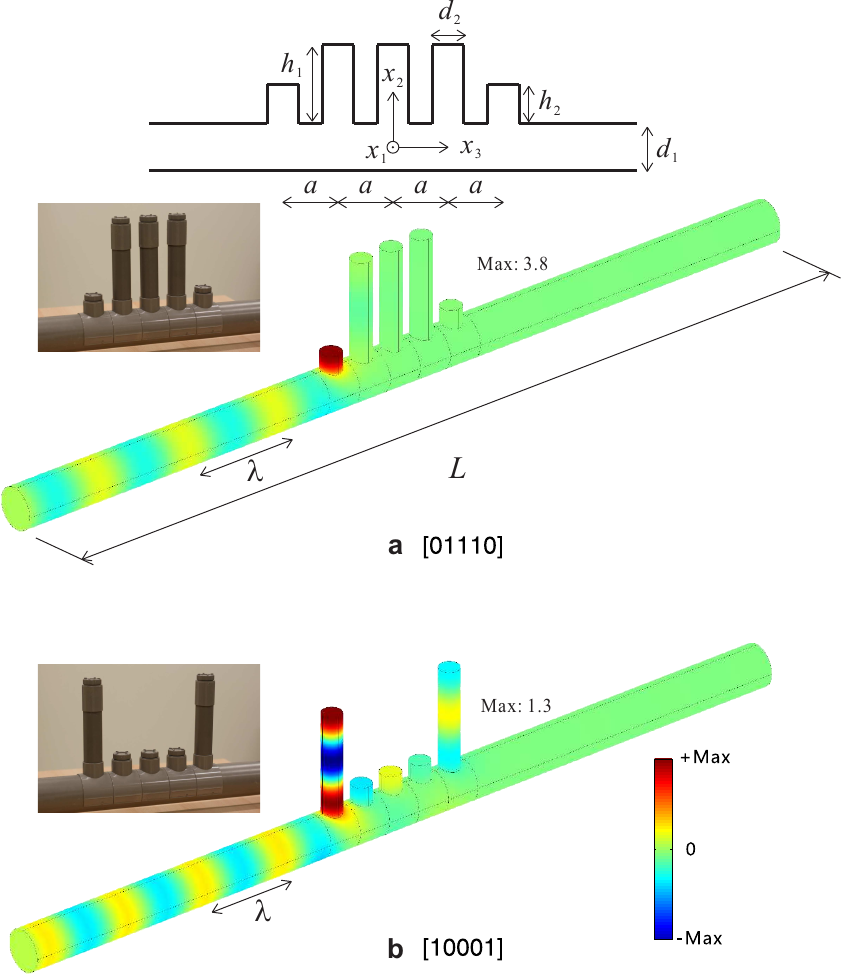}
\caption{
An acoustic system composed of a sequence of 5 short or long tubes grafted onto a waveguide.
The displayed sequences are denoted by (a) [01110] and (b) [10001], where 0 or 1 represent either a short or a long tube.
The distance between adjacent tubes is $a$.
Photographs of the actual samples are shown as insets.
Numerical simulations illustrating transmission cancellation are shown at the resonance frequency of (a) the short tube ($f=1456$ Hz) and (b) the long tube ($f=1638$ Hz).
The color-bar indicates the amplitude of the pressure field normalized to the maximum pressure (Max).
$\lambda$ is the wavelength at the considered frequency.
Dimensions in the experiment are $L=2$~m, $a=8$~cm, $d_1=10$~cm, $d_2=5$~cm, $h_1=24$~cm, and $h_2=4$~cm.}
\label{fig1}	
\end{figure}

The paper is organized as follows.
Experimental design and numerical methods used to obtain transmission spectra are described in Section \ref{sec2}.
Evanescent waves of the acoustic waveguide and their appearance at the grafting point are discussed in Section \ref{sec3}.
In Section \ref{sec4}, we turn our attention to aperiodic sequences with increasing separation between resonators.
Finally, our results are summarized in Section \ref{sec5}.

\section{Experimental and numerical methods}
\label{sec2}

For the purpose of demonstration, we consider possibly one of the simplest resonant systems: a one-dimensional sequence of air tubes as resonators grafted onto a waveguide \cite{fangNatMat2006,wangJPD2014}.
The principle of the experiment and some samples are shown in Fig. \ref{fig1}.
We use inexpensive cylindrical polyvinyl chloride (PVC) cylinders for both the waveguide and the resonators.
The waveguide supports only one propagating acoustic mode in air for frequencies under 2009~Hz; all experiments and computations are performed well below this cut-off frequency.
The dispersion relation of guided waves in a waveguide is specifically discussed in Section \ref{sec3.1}.

Resonators are chosen between tubes of two different lengths, either long ($h_1=24$ cm) or short ($h_2=4$ cm).
Such cylindrical tubes have a series of natural resonances that depend both on their length and on the boundary conditions at their ends.
With one end closed and the other opening inside a waveguide, natural resonances of a tube are up-shifted in frequency compared to the same tube closed at both ends; the frequency shift can be related to the appearance of evanescent waves at the grafting location \cite{wangJPD2014}.
The fundamental resonance of the grafted short tube is at 1456 Hz; the third resonance of the grafted long tube is at 1642 Hz.
By choosing close but different resonance frequencies, we wish to differentiate the strong coupling between tubes of equal length from the comparatively weaker coupling between tubes of unequal lengths (see the specific discussion in Section \ref{sec3.2}).
We consider aperiodic sequences of 5 grafted tubes that are either short ($0$) or long ($1$).
The distance between two adjacent tubes is set to $a=8$ cm, or about one third of the wavelength in air for all considered frequencies.
Transmission spectra are considered around the fundamental resonance frequency of the short tube, with sequences [00111], [01011], and [01110], i.e. for a progressive increase of the separation $l$ between short tubes.
For comparison, we consider also transmission spectra around the third resonance frequency of the long tube, with sequences [11000], [10100], and [10001].

\bigskip
Measurements were conducted using a sound card connected to a personal computer to generate Gaussian pulses with adjustable central frequency and bandwidth.
The pulses were played with a loudspeaker placed at one end of the waveguide.
The same sound card was used to sample the signals recorded with a microphone at the other end of the waveguide.
The sampling rate was chosen to be 384 kHz, which is amply sufficient to capture the relevant spectra. 
The acquisition duration was also chosen long enough to capture all signals exiting the waveguide with delay.
All experimental measurements are normalized against the measurement for a bare waveguide, i.e. with no resonator present, obtained with exactly the same Gaussian pulse.
Such a normalization procedure is used to smooth Fabry-Perot oscillations that appear from reflections at the ends of the 2-meter-long waveguide.
Though these oscillations are not completely canceled and can still appear in the normalized experimental transmission spectra, they have negligible influence on the experimental results reported in Section \ref{sec3}.
It is also noted that acoustic nonlinearities were verified to be negligible with our samples.
We checked this point by repeating our experiments at different signal levels.
Further details on the experimental measurement of isolated resonators and periodic sequences can be found in \cite{wangJPD2014}.

\bigskip
In order to evaluate numerically transmission through the samples, we used a three-dimensional time-harmonic finite element model of pressure wave propagation.
Examples of such simulations are shown in Fig. \ref{fig1}.
Numerical implementation was performed using the Acoustics module of the commercial software Comsol Multiphysics 3.5a.
The partial differential equation solved is the linear acoustics equation with pressure as the independent variable.
Material parameters used here are the mass density, $\rho=1.2041$ kg$.$m$^{-3}$, and the speed of sound, $c=342$~m.s$^{-1}$.
Moreover, sound attenuation is introduced in the calculation.
In the frequency range from 1360~Hz to 1720~Hz, the sound attenuation parameter is supposed to be proportional to frequency.
In comparison to the value for open air \cite{airloss}, the attenuation parameter we used is 3 times larger for a better match to experimental results.
This larger attenuation in our experiments may be due to additional losses arising from imperfections of the samples.
Continuity boundary conditions for pressure are imposed on all internal boundaries.
A radiation boundary condition including an incident harmonic plane wave with unit pressure ($p_0=1$ Pa) is imposed on the left entrance boundary (denoted as $\sigma_1$), and a radiation boundary condition canceling incoming waves is applied on the right exit boundary (denoted as $\sigma_2$).
All other external boundaries are set to the ``wall'' boundary condition, meaning that the normal velocity is zero on them, or equivalently that the normal derivative of pressure vanishes.
With this setting, we avoid reflections at the ends of the waveguide and the appearance of spurious FP oscillations in the simulated spectra.
Integration of the pressure field over the entrance and exit boundaries allows us to estimate the reflection and the transmission coefficients.
More precisely, the transmission coefficient, $t$, and the reflection coefficient, $r$, are evaluated as
\begin{equation}
t=\frac{\int_{\sigma_2} p dV}{A p_0},
\end{equation}
\begin{equation}
r=\frac{\int_{\sigma_1} p dV}{A p_0} -1,
\end{equation}
where $p$ is the pressure inside the waveguide and $A=\pi d_1^2/4$ is the surface area of the boundaries.
We let the angular frequency $\omega$ sweep the frequency range of interest in order to obtain spectra.
Note that these formulas work if there is only one propagating guided mode and  boundaries $\sigma_1$ and $\sigma_2$ are in the far field of the resonators, so that evanescent waves can be assumed to have decayed sufficiently to be unnoticeable at those boundaries.

\section{Evanescent waves originating from a grafting point}
\label{sec3}

\begin{figure}[!t]
\centering
\includegraphics{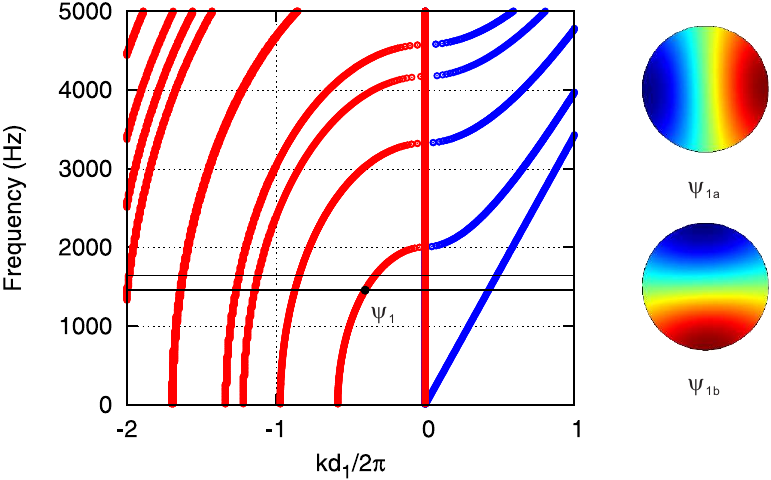}
\caption{Dispersion relation of the guided modes of an acoustic waveguide of diameter $d_1$, obtained by solving for the square of the wavenumber, $k^2$, as a function of frequency.
Both real (propagating, blue color) and imaginary (evanescent, red color) branches of the dispersion relation are shown.
Note that the branches with opposite sign of $k$ exist as well.
The cut-off frequency of the first guided mode is 2009 Hz.
The two black lines mark the two resonances considered in this work.
Pressure distributions for the two degenerate modes at the marked point on the first imaginary branch are shown on the right side.}
\label{fig2}
\end{figure}

\subsection{Guided evanescent waves}
\label{sec3.1}

The dispersion relation of guided waves in a homogeneous acoustic waveguide are obtained by solving the acoustic wave equation
\begin{equation}
\frac{\partial ^2 p }{\partial x_1^2} + \frac{\partial ^2 p }{\partial x_2^2} + \frac{\partial ^2 p }{\partial x_3^2}= - \frac{\omega ^2}{c^2} p.
\label{eq3}
\end{equation}
under the assumption that guided waves can be expressed as 
\begin{equation}
p = \psi(x_1,x_2) \exp(-\imath k x_3),
\label{eq4}
\end{equation}
where $\psi$ is the transverse modal distribution.
Substituting Eq. (\ref{eq4}) into Eq. (\ref{eq3}), we get
\begin{equation}
\frac{\partial ^2 \psi }{\partial x_1^2} + \frac{\partial ^2 \psi }{\partial x_2^2} + \frac{\omega ^2}{c^2} \psi = k^2 \psi ,
\label{eq5}
\end{equation}
where $k$ is the wavenumber of the guided waves.

The complex dispersion relation of guided waves shown in Fig. \ref{fig2} is obtained by sweeping the frequency $f=\omega/(2 \pi)$ in the range of interest and obtaining $k^2$ as the eigenvalue and $\psi$ as the eigenvector in Eq. \eqref{eq5}.
At a given frequency, the wavenumber is either real or pure imaginary, corresponding to either propagating or evanescent guided modes.
The fundamental resonance frequency of the short grafted tube and the third resonance frequency of the grafted long tube are indicated in Fig. \ref{fig2} by the two horizontal black lines.
Around these frequencies, only the fundamental guided mode can propagate in the waveguide and all other modes are evanescent.
The fundamental guided mode is non dispersive and propagates at celerity $c$.
For a homogeneous waveguide, the transverse modal function $\psi_0 (x_1,x_2)$ is simply a constant.
All other higher-order guided modes are evanescent and dispersive: their wavenumbers are imaginary-valued functions of frequency and their amplitude decreases exponentially along the propagation direction.
Owing to the symmetry of the waveguide, the first evanescent dispersion branch supports two independent and frequency-degenerate modes, as shown in Fig. \ref{fig2}.
These modes $\psi_{1a}$ and $\psi_{1b}$ are mutually orthogonal.
Their transverse pressure distributions are typically representative of half-wavelength standing wave patterns.

\subsection{Single grafted tube}
\label{sec3.2}

\begin{figure}[!t]
\centering
\includegraphics{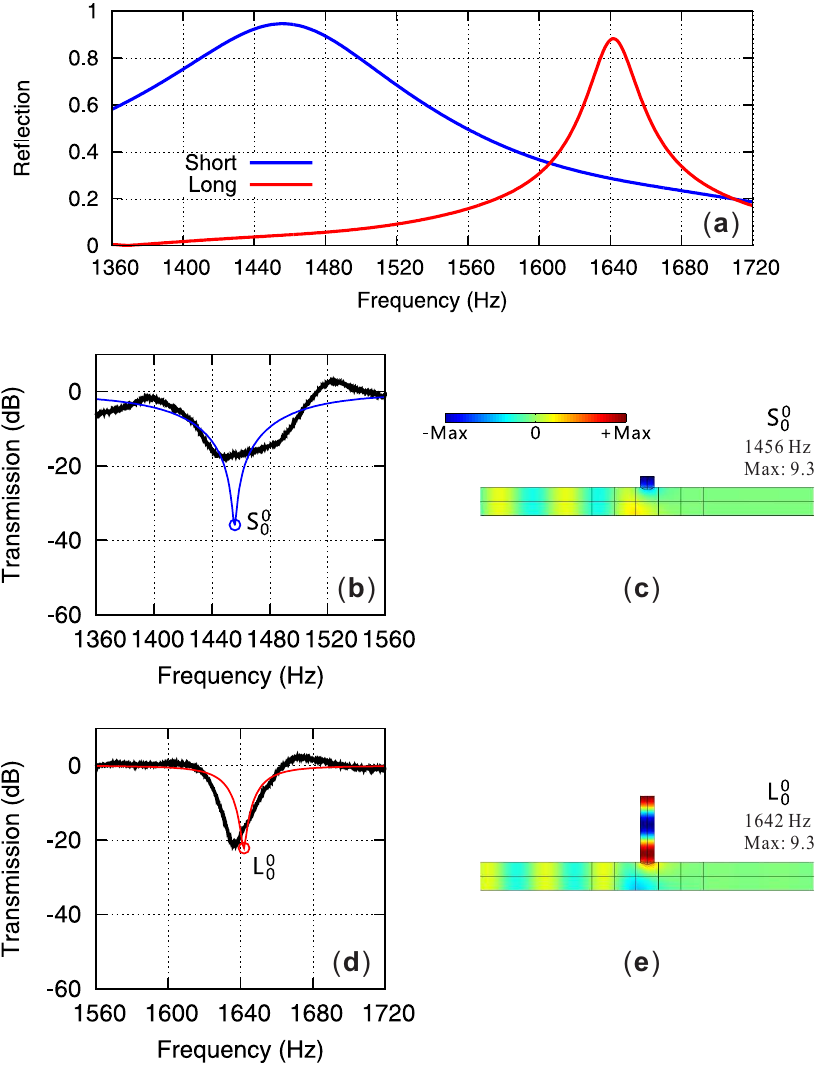}
\caption{
(a) Numerical reflection coefficients for the single short (blue line) or the single long (red line) tube grafted onto a waveguide.
(b) Experimental and numerical transmission spectra for the single short tube grafted onto a waveguide. 
(c) Pressure distribution field at the resonant frequency for the short tube.
The color bar represents the normalized pressure with respect to the maximum pressure (Max) at the corresponding frequency.
(d)-(e) are similar to (b)-(c) for the case of the single long tube.}
\label{fig3}
\end{figure}

Considered separately, a single grafted tube mostly acts as a localized and strongly dispersive mirror having a reflection coefficient peaking almost at unity at the resonance frequency.
Actually, unit reflection would be obtained in the absence of loss.
Fig. \ref{fig3}(a) shows the computed reflection coefficient for a single short or long tube in the frequency range of interest.
At the fundamental resonance ($f=1456$~Hz) of the short tube, the reflection coefficient of the long tube is quite small ($|r| \approx 0.1$).
This implies that the long tube has small impact on transmission at the fundamental resonance of the short tube.
In contrast, at the third resonance ($f=1642$~Hz) of the long tube, the reflection coefficient on the short tube retains a significant value ($|r| \approx 0.3$).
Similarly, this implies that the short tube could have noticeable impact on transmission at the resonance of the long tube.
Thus, when one type of tube is at resonance, the other type is not but its presence still cannot be completely neglected.

Figs. \ref{fig3}(b) and \ref{fig3}(d) present experimental measurements and numerical simulations for the single short or long tubes grafted onto a waveguide.
A transmission dip following the standard Lorentzian profile is introduced at the resonance frequency, as should be expected \cite{wangJPD2014}.
The pressure distributions at resonance are shown for the single short tube in Fig. \ref{fig3}(c) and for the single long tube in Fig. \ref{fig3}(e).
Both show that a standing wave is formed to the left of the tube due to the interference of the incident propagating guided wave and the reflected propagating guided wave.
Evanescent guided waves are clearly formed at the grafting points.
They appear because the translational symmetry of the waveguide is broken at the grafting position and they ensure the continuity of pressure before and after it.

Considering only the fundamental propagating guided mode and the first evanescent guided mode, the total pressure field inside the waveguide can be expressed as
\begin{equation}
p = \left \{
\begin{array}{ccc}
\psi_0 e^{-\imath k x_3} + r \psi_0 e^{\imath k x_3} + s \psi_{1b} e^{\alpha x_3} & \mathrm{if} & x_3<0, \\
t \psi_0 e^{- \imath k x_3} + s \psi_{1b} e^{- \alpha x_3} & \mathrm{if} & x_3>0,
\end{array}
\right .
\label{eq6}
\end{equation}
where $k=\omega/c$, $\alpha=|\Im(k_1(\omega))|$ is the modulus of the imaginary part of the wavenumber for the first evanescent guided mode, and $s$ is the evanescence coefficient \cite{wangJPD2014}.
Of the first two evanescent guided modes, only $\psi_{1b}$ is excited because of the position of the grafting point on the waveguide.
At a resonance frequency, the reflection and transmission coefficients are $|r| \approx 1$ and $|t| \approx 0$, respectively.
Then Eq. (\ref{eq6}) can be rewritten as
\begin{equation}
p \approx \left \{
\begin{array}{ccc}
\psi_0 (e^{-\imath k x_3} + r e^{\imath k x_3}) + s \psi_{1b} e^{\alpha x_3} & \mathrm{if} & x_3<0, \\
s \psi_{1b} e^{- \alpha x_3} & \mathrm{if} & x_3>0.
\end{array}
\right .
\label{eq7}
\end{equation}
Since the amplitude of the transmitted guided wave is almost zero, the extent of evanescent waves along the direction of propagation can be visualized clearly on the right side of the grafting point.
Moreover, the evanescent contribution in Eq. (\ref{eq7}) decreases rapidly to zero with increasing distance from the grafting point, see Table \ref{table1}.
Therefore, the standing wave pattern formed on the left side of the resonator as well as the canceling of transmission on its right side can be clearly observed in the far field of the pressure distributions of Figs. \ref{fig3}(c) and \ref{fig3}(e).

\begin{table}[!t]
\centering
\caption{Amplitude decay of evanescent guided waves $\exp(- \alpha n a)$ as a function of the imaginary part of the wavenumber ($\alpha$) and of the propagation distance $na$.}
\label{table1}
\begin{tabular}{p{30mm}p{10mm}p{10mm}p{10mm}p{10mm}} \toprule[2pt]
& \multicolumn{4}{c}{$n$} \\ \cline{2-5}
$\alpha$ (m$^{-1}$) & 0 & 1 & 2 & 4 \\ \midrule[1pt]
21.1 @ 1456 Hz & 1 & 0.18 & 0.034 & 0.0012\\
25.3 @ 1642 Hz & 1 & 0.13 & 0.017 & 0.0003\\ \bottomrule[2pt]
\end{tabular}\\
\end{table}

\section{Aperiodic sequences of resonators}
\label{sec4}

\begin{figure}[!t]
\centering
\includegraphics{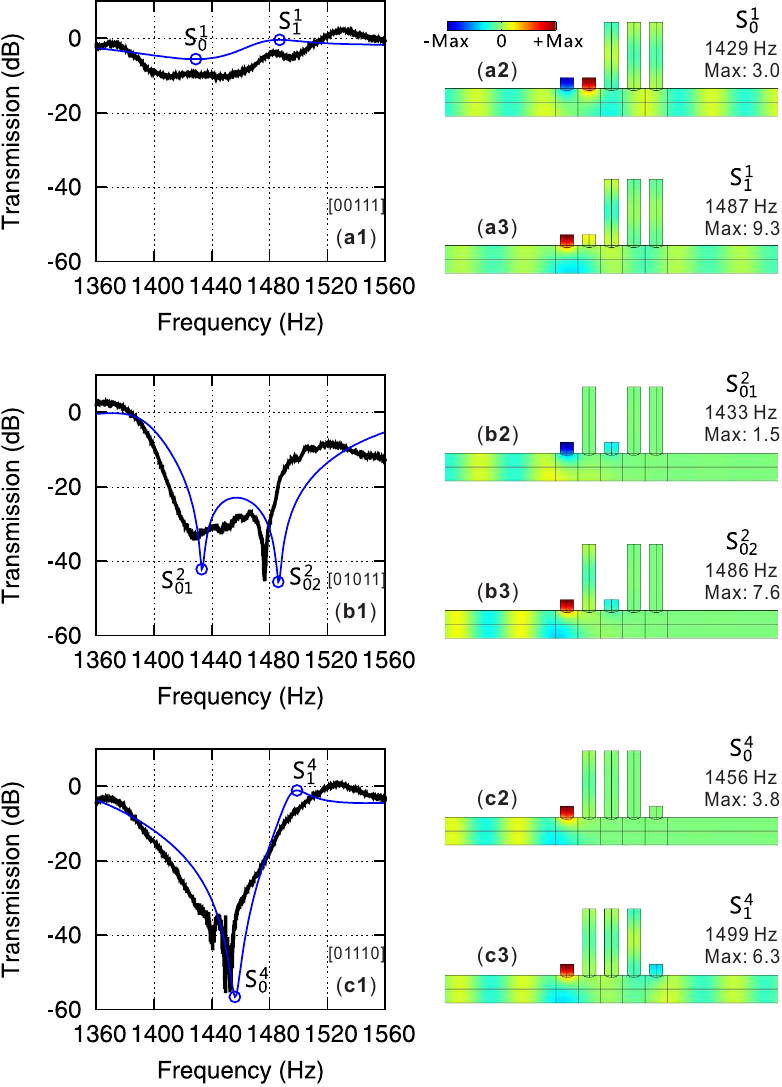}
\caption{Transmission through a sequence of 5 short or long tubes grafted onto a waveguide with increasing separation between short tubes for (a) $l=a$, (b) $l=2a$, and (c) $l=4a$.
Other dimensions are defined in Fig. \ref{fig1}.
Experimental (thick black line) computed (thin blue line) transmission spectra around the fundamental resonance frequency of the short tube.
Remarkable frequencies are labeled S$^\alpha_{\beta \gamma}$, with $\alpha=l/a$ indicating the normalized separation between the two long tubes and $\beta$ indicating either a transmission dip (0) or unit (1).
If there are two transmission dips, they are identified using $\gamma=1,2$.
Normalized pressure distributions at the remarkable frequencies are shown on the right side.
For simplicity, we display only vertical cross-sections going through the symmetry axis of the waveguide.
In each case, the frequency of the incident harmonic guided waves is shown on top.
The color scale is for pressure normalized with respect to the maximum pressure (Max).
}
\label{fig4}
\end{figure}

\begin{figure}[!t]
\centering
\includegraphics{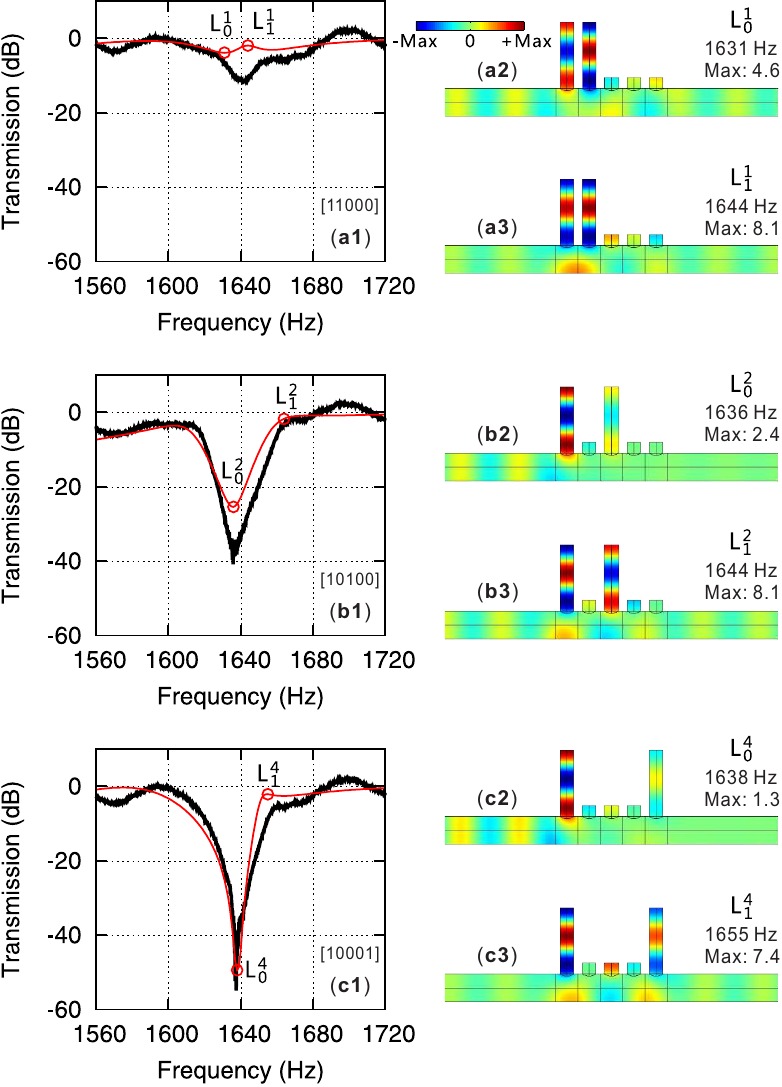}
\caption{Transmission through a sequence of 5 short or long tubes grafted onto a waveguide with increasing separation between long tubes for (a) $l=a$, (b) $l=2a$, and (c) $l=4a$.
Other dimensions are defined in Fig. \ref{fig1}.
Experimental (thick black line) computed (thin red line) transmission spectra around the fundamental resonance frequency of the short tube.
Remarkable frequencies are labeled L$^\alpha_{\beta}$, with $\alpha=l/a$ indicating the normalized separation between the two long tubes and $\beta$ indicating either a transmission dip (0) or unit (1).
Normalized pressure distributions at the remarkable frequencies are shown on the right side.
For simplicity, we display only vertical cross-sections going through the symmetry axis of the waveguide.
In each case, the frequency of the incident harmonic guided waves is shown on top.
The color scale is for pressure normalized with respect to the maximum pressure (Max).
}
\label{fig5}
\end{figure}

From this point on we discuss measurements and numerical simulations for sequences of 5 grafted tubes.
Fig. \ref{fig4} reports the transmission spectra in the frequency range of the fundamental resonance of the short tube, considering aperiodic sequences [00111], [01011], and [01110]; it also shows the pressure distribution at remarkable frequencies for which transmission minima and maxima are observed.
Those remarkable frequencies are labeled S$^\alpha_{\beta \gamma}$, with ``S'' referring to the short tube and with $\alpha=l/a$ indicating the normalized separation between the two short tubes and $\beta$ indicating either a transmission dip (0) or unit (1).
Fig. \ref{fig5} contain similar information in the frequency range of the third resonance of the long tube, for aperiodic sequences [11000], [10100], and [10001].
In this case, remarkable frequencies are labeled L$^\alpha_{\beta}$, with ``L'' referring to the long tube.

\subsection{Near-field coupling regime}
\label{sec4.1}

Let us now consider the case of sequence [00111] with two adjacent resonant short tubes (case $l=a$ in Fig.~\ref{fig4}(a)).
Amazingly, the transmission dip disappears almost completely and transmission to the propagating guided wave is close to unity for all frequencies.
It does not mean, however, that the two short tubes are out of resonance, as can be checked from the pressure distributions at frequencies S$_0^1$ and S$_1^1$ in Fig. \ref{fig4}(a).
Instead, each resonator strongly feels the evanescent waves attached to the other resonator and a strong near-field coupling results.
Furthermore, it was checked by reproducing the numerical simulation without the long tubes that their presence almost does not change the transmission.
This fact is consistent with the result in Fig. \ref{fig3} that the long tube has a small reflection coefficient around the fundamental resonance of the short tube.
Interchanging the role of short and long tubes with sequence [11000] (Fig. \ref{fig5}(a)), we note that the general trends remain the same.

When the separation between the two short tubes is enlarged to $l=2a$ with sequence [01011], a system of two transmission dips results in Fig. \ref{fig4}(b).
Observing the pressure distributions at frequencies S$_{01}^2$ and S$_{02}^2$, however, the first short tube is more strongly excited and plays a dominant role in the cancellation of transmission.
The difference between both states seems to be that pressure in the second short tube oscillates with the same or the reverse phase compared to the first.
In the case of the [10100] sequence around the resonance of the long tube in Fig. \ref{fig5}(b), there is only a single transmission dip.
It was checked numerically that the simple presence or absence of the off-resonance short tubes leads to either a single dip or a system of two dips in the transmission spectrum.
Again, this fact is consistent with the result in Fig. \ref{fig3} that the short tube has a relatively significant reflection coefficient around the third resonance of the long tube.
These observations are a strong indication that a small perturbation of the environment of the long tubes, through weak near-field coupling, can lead to significant changes in the transmission.

\subsection{Locally-resonant Fabry-Perot interferometers}
\label{sec4.2}

When considering a larger separation ($l = 4a$) between short tubes, such as for sequences [01110], it is expected that the near-field coupling that is due to evanescent waves gets smaller and eventually vanishes exponentially, in accordance with Table \ref{table1}.
The transmission spectra in Fig. \ref{fig4}(c) start to assume an asymmetric Fano-like shape \cite{goffauxPRL2002, miroshnichenkoRMP2010} instead of the simpler Lorentzian shape observed for smaller separations.
More precisely, this shape is characterized by the simultaneous presence of a transmission dip centered near the resonance frequency of the single tube and of one or more transmission units.
At the transmission dip S$_0^4$, it can be observed in Fig. \ref{fig4}(c2) that acoustic waves localize on the first tube while the second tube remains almost idle.
This property was checked to remain true for larger separations.
At the transmission unit S$_1^4$, both short tubes are rather strongly excited in Fig. \ref{fig4}(c3) and a standing wave establishes between them.
Similar observations can be made regarding the transmission spectrum, the transmission dip L$_0^4$, and the transmission unit L$_1^4$ in Fig.~\ref{fig5}(c).

\begin{figure}[!t]
\centering
\includegraphics{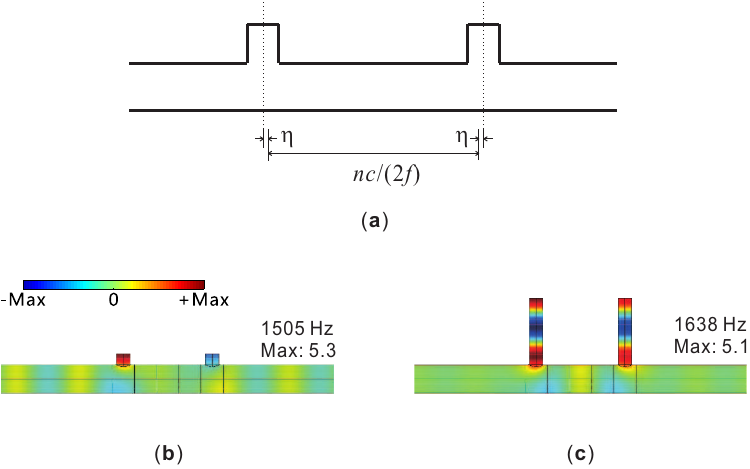}
\caption{
(a) Schematic of a locally resonant Fabry-Perot interferometer formed by localized reflections.
Reflections on the resonators are effectively shifted by distance $\eta$ from the grafting center.
Fabry-Perot resonances are formed when the separation $l$ between resonators is the sum of an integer number of half-wavelengths plus twice distance $\eta$.
(b,c) Normalized pressure distributions at resonance for locally resonant Fabry-Perot interferometers in case of short or long tubes, for $l=4a$.
The color scale is for pressure normalized with respect to the maximum pressure (Max).	}
\label{fig6}
\end{figure}

We interpret transmission units as caused by locally resonant FP interferometers formed by localized reflections on the resonators.
Indeed, since the resonating tubes reflect efficiently propagating guided waves, a system of standing waves can be formed between two successive resonant tubes.
According to this picture, the number of transmission units must increase proportionally to the separation between the two resonating tubes.
Their frequencies are given by the condition that the effective round-trip FP cavity length is an integral number $n$ of wavelengths, leading to the equation
\begin{equation}
l = N a = n \frac{\lambda}{2} + 2\eta = n \frac{c}{2 f} + 2\eta,
\label{eq8}
\end{equation}
where $\eta$ is an effective distance accounting for the fact that reflection does not occur exactly at the center of the tubes, because of the excitation of evanescent waves.
This effective distance is a frequency-dependent quantity related to the details of the grafting geometry and is only obtained numerically in this work, as illustrated in Fig. \ref{fig6}.

According to Eq. \eqref{eq8}, the resonances of locally resonant FP interferometers are given by parameters $n$ and $\eta$, for a given separation $l=N a$.
For instance, the parameters defining the frequencies of FP resonances in Fig.~\ref{fig6}(b) are $n=3$ and $\eta=-0.13a$;
parameters defining the frequencies of FP resonances in Fig.~\ref{fig6}(c) are $n=3$ and $\eta=0.04a$.

The influence on transmission units of off-resonance tubes present inside a locally resonant FP interferometer depends on the type of the additional tubes.
For resonating short tubes, the presence or absence of off-resonance long tubes does not significantly affect the transmission: in Fig.~\ref{fig4}(c), the frequency of the transmission unit is almost the same as the frequency in Fig.~\ref{fig6}(b).
Conversely, for resonating long tubes, the presence or absence of off-resonance short tubes affects the transmission: in Fig. \ref{fig5}(c), the frequency of the transmission unit moves to the other side of the transmission dip compared to Fig.~\ref{fig6}(c).
In a practical system based on resonators grafted on a waveguide, the transmission could thus be tuned efficiently by adjusting the properties of additional off-resonance resonators within a locally resonant FP interferometer.

\subsection{Two identical resonators grafted onto the waveguide with a large separation}
\label{sec4.3}

\begin{figure*}[!t]
\centering
\includegraphics[width=140mm]{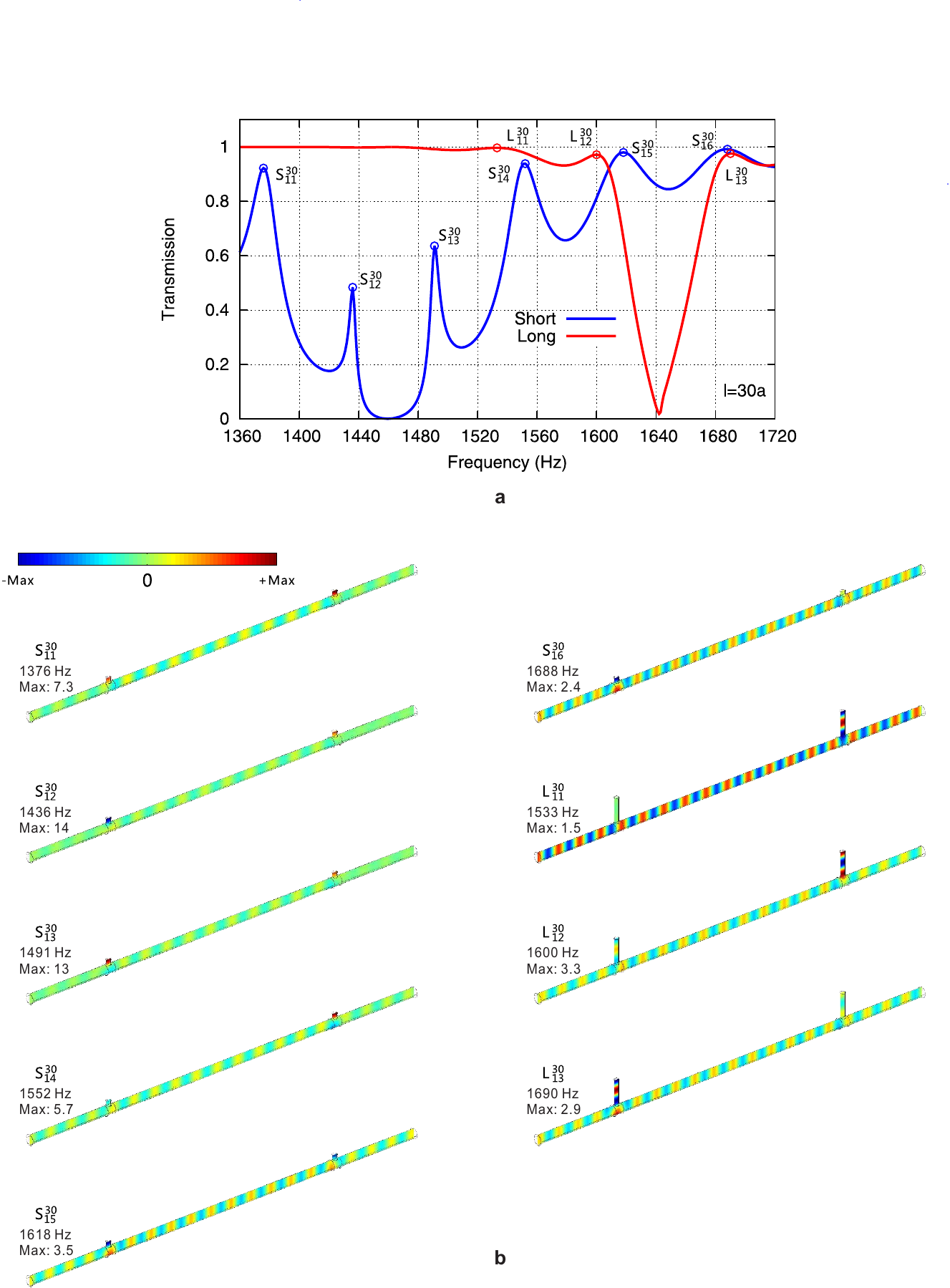}
\caption{Case of two resonators grafted onto a waveguide at distance $l=30a$ ($l \sim 10 \lambda$).
(a) Transmission spectra are shown for 2 short tubes (solid blue line) or 2 long tubes (solid red line).
(b) Normalized pressure distributions are shown at the remarkable frequencies indicated in (a).
The color scale is for pressure normalized with respect to the maximum pressure (Max).	}
\label{fig7}
\end{figure*}

According to the analysis in the previous section, two identical tubes grafted at an arbitrary distance $l$ along a waveguide create a locally resonant FP interferometer.
The validity of this concept can be checked numerically for a rather large separation of the tubes.
In this case, $n$ in Eq. (\ref{eq8}) is large and we expect that the sequence of FP resonances will be given by the approximate formula
\begin{equation}
f_n \approx n \frac{c}{2 l} ,
\label{eq9}
\end{equation}
or a sequence of harmonics of a fundamental frequency.

Fig. \ref{fig7} shows transmission spectra in the case that $l=30 a$.
This distance is $l=2.4$ m, about 10 times larger than the wavelength ($\lambda \approx 0.23$ m).
In this case, the two resonators could naively be supposed to be insensitive to one another.
In fact, coupling of their evanescent fields can safely be ignored, but interference of reflected guided waves can still occur.
Fig. \ref{fig7}(a) shows transmission spectra for pairs of short or long tubes.
The series of transmission units is seen to be given with a good approximation by formula (\ref{eq9}).
The sequence of states S$^{30}_{11}$ to S$^{30}_{16}$ are actually associated with the values $n=19$ to 24.
The sequence of states L$^{30}_{11}$ to L$^{30}_{13}$ also correspond to $n=22$ to 24.
The pressure distributions for all these states are plotted in Fig. \ref{fig7}(b).
FP oscillations with enhanced amplitude can clearly be seen to extend between the two resonating tubes.

As a note, a coefficient of finesse indicative of the width of the transmission peaks can be defined as
\begin{equation}
F = \frac{4 |r_0|^2}{(1 - |r_0|^2)^2}
\end{equation}
where $r_0$ is the reflection coefficient on one resonator at the considered frequency.
From this expression, it is clearly seen that high finesses can only be obtained as $|r_0|$ tends to unity.
Locally-resonant FP interferometers can thus be made highly selective in frequency as a result of the strong dependence of the reflection coefficient with frequency.
It should be noted, however, that if the separation between tubes is tuned to maximize the finesse, then a transmission unit will always be very close to a transmission zero, since both conditions amount to choosing $|r_0| \approx 1$.

\section{Conclusion}
\label{sec5}

We have observed that longitudinal coupling between resonant elements in an acoustic system can be very strong in the near-field regime and can be exploited to tune the transmission.
Indeed, we formed sequences of resonators with almost no attenuation or with a strong attenuation over a rather wide frequency bandwidth.
We observed strong coupling between resonators with the same resonance frequency and weak coupling between resonators with different but close resonance frequencies.
Moreover, when the separation between two identical resonators was larger than about half a wavelength, we observed the additional appearance of transmission units arising from the formation of locally resonant Fabry-Perot interferometers.

Adopting a wider view, we remark that models of metamaterials based on homogenization techniques or effective parameters generally consider propagating waves in the long wavelength regime.
They thus implicitly neglect the possible formation of evanescent waves that can extend beyond the limits of a unit-cell and couple embedded resonators along the direction of propagation.
In such a case, longitudinal near-field coupling should not be ignored and can provide new degrees of freedom for wave propagation control at the sub-wavelength scale.

The acoustic system we have considered is one-dimensional and is easily reconfigurable \cite{driscollScience2009, taoPRL2009, ouNanoL2011, caleapPNAS2014} just by adjusting the length of the resonators and the order of their sequence.
Longitudinal near-field coupling should be observable in two- and three-dimensional acoustic systems as well, since waves that are evanescent in the direction of propagation must also appear at the grafting positions of the resonators.
Beyond acoustic waves, the same considerations extend naturally to elastic waves, though their vector character clearly would need to be taken into account.

\section*{Acknowledgments}

We thank Dr. Sarah Benchabane and Pr. Yue-Sheng Wang for fruitful discussions.
Photographs of samples are courtesy of R\'emi Meyer.
Financial support by the Labex ACTION program (Contract No. ANR-11-LABX-0001-01) is gratefully acknowledged.


\bibliographystyle{unsrt}
\bibliography{nearfield}

\begin{thebibliography}{10}

\bibitem{liuSCIENCE2000}
Z.Y. Liu, X.~Zhang, Y.~Mao, Y.Y. Zhu, Z.~Yang, C.T. Chan, and P.~Sheng.
\newblock Locally resonant sonic materials.
\newblock {\em Science}, 289:1734--1736, 2000.

\bibitem{laiNatMater2011}
Y.~Lai, Y.~Wu, P.~Sheng, and Z.~Zhang.
\newblock Hybrid elastic solids.
\newblock {\em Nature Materials}, 10:620--624, 2011.

\bibitem{zhuNatC2014}
R.~Zhu, X.N. Liu, G.K. Hu, C.T. Sun, and G.L. Huang.
\newblock Negative refraction of elastic waves at the deep-subwavelength scale
  in a single-phase metamaterial.
\newblock {\em Nature Communications}, 5:5510, 2014.

\bibitem{wang2011JAP}
Y.~F. Wang, Y.~S. Wang, and X.~X. Su.
\newblock Large bandgaps of two-dimensional phononic crystals with cross-like
  holes.
\newblock {\em Journal of Applied Physics}, 110:113520, 2011.

\bibitem{wang2004PRL}
G.~Wang, X.~Wen, J.~Wen, L.~Shao, and Y.~Liu.
\newblock Two-dimensional locally resonant phononic crystals with binary
  structures.
\newblock {\em Physical Review Letters}, 93:154302, 2004.

\bibitem{elford2011jasa}
D.P. Elford, L.~Chalmers, F.V. Kusmartsev, and G.M. Swallowe.
\newblock Matryoshka locally resonant sonic crystal.
\newblock {\em Journal of the Acoustical Society of America}, 130:2746--2755,
  2011.

\bibitem{yilmaz2007prb}
C.~Yilmaz, G.M. Hulbert, and N.~Kikuchi.
\newblock Phononic band gaps induced by inertial amplification in periodic
  media.
\newblock {\em Physical Review B}, 76:054309, 2007.

\bibitem{ma2013OE}
T.~X. Ma, Y.~S. Wang, Y.~F. Wang, and X.~X. Su.
\newblock Three-dimensional dielectric phoxonic crystals with network topology.
\newblock {\em Optics Express}, 21:2727--2732, 2013.

\bibitem{pennec2008PRB}
Y.~Pennec, B.~Djafari-Rouhani, H.~Larabi, J.O. Vasseur, and A.C.
  Hladky-Hennion.
\newblock Low-frequency gaps in a phononic crystal constituted of cylindrical
  dots deposited on a thin homogeneous plate.
\newblock {\em Physical Review B}, 78:104105, 2008.

\bibitem{achaoui2011PRB}
Y.~Achaoui, A.~Khelif, S.~Benchabane, L.~Robert, and V.~Laude.
\newblock Experimental observation of locally-resonant and {B}ragg band gaps
  for surface guided waves in a phononic crystal of pillars.
\newblock {\em Physical Review B}, 83:104201, 2011.

\bibitem{fangNatMat2006}
N.~Fang, D.~Xi, J.~Xu, M.~Ambati, W.~Srituravanich, C.~Sun, and X.~Zhang.
\newblock Ultrasonic metamaterials with negative modulus.
\newblock {\em Nature Materials}, 5:452--456, 2006.

\bibitem{lemoult2011PRL}
F.~Lemoult, M.~Fink, and G.~Lerosey.
\newblock Acoustic resonators for far-field control of sound on a subwavelength
  scale.
\newblock {\em Physical Review Letters}, 107:064301, 2011.

\bibitem{husseinAMR2014}
M.I. Hussein, M.J. Leamy, and M.~Ruzzene.
\newblock Dynamics of phononic material and structures: {H}istorical origins,
  recent progess, and future outlook.
\newblock {\em Applied Mechanics Reviews}, 66:040802, 2014.

\bibitem{fuAFM2011}
Y.H. Fu, A.Q. Liu, W.M. Zhu, X.M. Zhang, D.P. Tsai, J.B. Zhang, T.~Mei, J.F.
  Tao, H.C. Guo, X.H. Zhang, et~al.
\newblock A micromachined reconfigurable metamaterial via reconfiguration of
  asymmetric split-ring resonators.
\newblock {\em Advanced Functional Materials}, 21:3589--3594, 2011.

\bibitem{liuNatPhoto2009}
N.~Liu, H.~Liu, S.~Zhu, and H.~Giessen.
\newblock Stereometamaterials.
\newblock {\em Nature Photonics}, 3:157--162, 2009.

\bibitem{powellPRB2011}
D.A. Powell, K.~Hannam, I.V. Shadrivov, and Y.S. Kivshar.
\newblock Near-field interaction of twisted split-ring resonators.
\newblock {\em Physical Review B}, 83:235420, 2011.

\bibitem{wangJPD2014}
Y.~F. Wang, V.~Laude, and Y.~S. Wang.
\newblock Coupling of evanescent and propagating guided modes in locally
  resonant phononic crystals.
\newblock {\em Journal of Physics D: Applied Physics}, 47:475502, 2014.

\bibitem{airloss}
Sound attenuation coefficient. {A}vailable online:
  \url{https://en.wikibooks.org/wiki/Engineering_Acoustics/Outdoor_Sound_Propagation}.

\bibitem{goffauxPRL2002}
C.~Goffaux, J.~S{\'a}nchez-Dehesa, A.L. Yeyati, Ph. Lambin, A.~Khelif, J.O.
  Vasseur, and B.~Djafari-Rouhani.
\newblock Evidence of {F}ano-like interference phenomena in locally resonant
  materials.
\newblock {\em Physical Review Letters}, 88:225502, 2002.

\bibitem{miroshnichenkoRMP2010}
A.E. Miroshnichenko, S.~Flach, and Y.S. Kivshar.
\newblock Fano resonances in nanoscale structures.
\newblock {\em Reviews of Modern Physics}, 82:2257--2298, 2010.

\bibitem{driscollScience2009}
T.~Driscoll, H.T. Kim, B.G. Chae, B.J. Kim, Y.W. Lee, N.M. Jokerst, S.~Palit,
  D.R. Smith, M.~Di~Ventra, and D.N. Basov.
\newblock Memory metamaterials.
\newblock {\em Science}, 325:1518--1521, 2009.

\bibitem{taoPRL2009}
H.~Tao, A.C. Strikwerda, K.~Fan, W.J. Padilla, X.~Zhang, and R.D. Averitt.
\newblock Reconfigurable terahertz metamaterials.
\newblock {\em Physical Review Letters}, 103:147401, 2009.

\bibitem{ouNanoL2011}
J.Y. Ou, E.~Plum, L.~Jiang, and N.I. Zheludev.
\newblock Reconfigurable photonic metamaterials.
\newblock {\em Nano Letters}, 11:2142--2144, 2011.

\bibitem{caleapPNAS2014}
M.~Caleap and B.W. Drinkwater.
\newblock Acoustically trapped colloidal crystals that are reconfigurable in
  real time.
\newblock {\em Proceedings of the National Academy of Sciences of the United
  States}, 111:6226--6230, 2014.

\end{thebibliography}

\end{document}